\documentclass{article}
\usepackage{spconf,amsmath,graphicx,hyperref}
\usepackage{multirow}
\usepackage{array}
\usepackage{pifont}
\usepackage{xcolor}
\usepackage{hyperref}
\usepackage[hyphenbreaks]{breakurl}

\title{CompSpoof: A Dataset and Joint Learning Framework for Component-Level Audio Anti-spoofing Countermeasures}

\name{Xueping Zhang$^{1}$,   Yechen Wang$^{2}$, Linxi Li$^{2}$, Liwei Jin$^{2}$, Ming Li$^{1}$ \thanks{Corresponding Author: Ming Li, ming.li369@dukekunshan.edu.cn}}

\address{$^{1}$Suzhou Municipal Key Laboratory of Multimodal Intelligent Systems,\\ Digital Innovation Research Center,
Duke Kunshan University, Kunshan, China \\ $^{2}$OfSpectrum, Inc., Los Angeles, USA}

\begin{document}

\maketitle

\begin{abstract}
Component-level audio Spoofing (CompSpoof) targets a new form of audio manipulation where only specific components of a signal, such as speech or environmental sound, are forged or substituted while other components remain genuine. Existing anti-spoofing datasets and methods treat an utterance or a segment as entirely bona fide or entirely spoofed, and thus cannot accurately detect component-level spoofing. To address this, we construct a new dataset, CompSpoof, covering multiple combinations of bona fide and spoofed speech and environmental sound. We further propose a separation-enhanced joint learning framework that separates audio components apart and applies anti-spoofing models to each one. Joint learning is employed, preserving information relevant for detection. Extensive experiments demonstrate that our method outperforms the baseline, highlighting the necessity of separate components and the importance of detecting spoofing for each component separately. Datasets and code are available at: \url{https://github.com/XuepingZhang/CompSpoof}.
\end{abstract}
\section{Introduction}
\label{sec:intro}

\begin{figure*}[htbp]
\centerline{\includegraphics[width=1\textwidth]{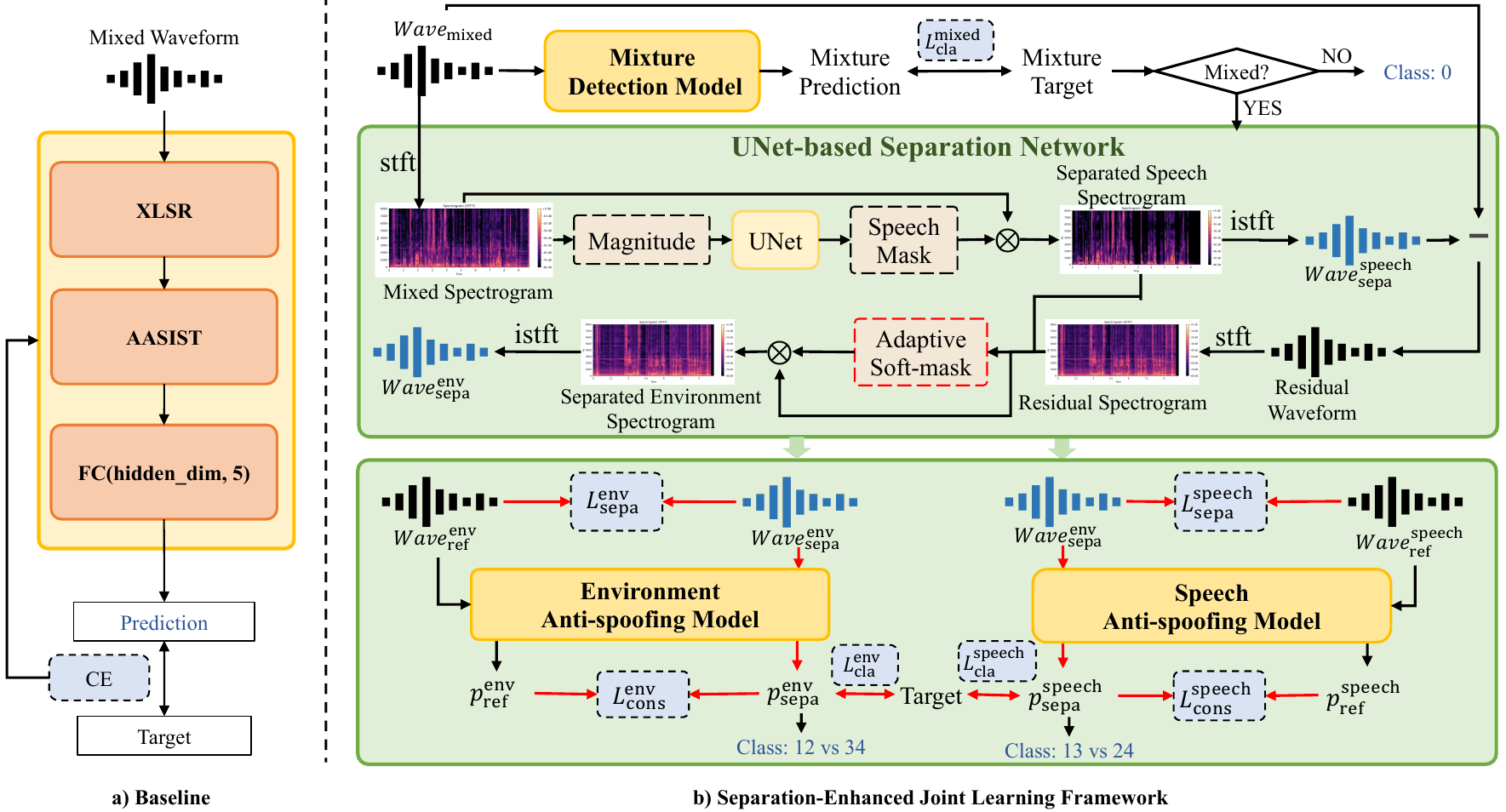}}
\caption{Overview of the baseline and proposed separation-enhanced joint learning framework. `` \textcolor{red}{$\rightarrow$} ' illustrates the joint learning data flow between the separation and anti-spoofing models.}
\label{main}
\end{figure*}

Component-level audio anti-spoofing addresses a new type of audio manipulation where only specific components of a signal are forged or substituted while the rest remains authentic. Unlike conventional spoofing attacks \cite{tts, vc} that generate or convert an entire utterance or segment along the time axis, component spoofing operates at a finer granularity: for instance, the speech content may be substituted with a synthetic voice while keeping the genuine environment noise, or conversely, the original speech may be preserved while the environment sound is generated or substituted. Such component manipulations are more complex to detect because they can slip through systems made for whole-utterance or time-domain partial spoofing \cite{frame_deepfake, partial}, and also sound more real to human listeners.

Over the past decade, the ASVspoof \cite{asvspoof21, asvspoof5} and other related datasets \cite{for, ITW, ai4tt, EnvSDD} have driven significant progress in audio anti-spoofing research. Current systems \cite{aasist, Wav2vec2, continual,sys_few} typically formulate spoof detection as binary classification between bona fide and spoofed utterances. Existing anti-spoofing methods \cite{Wav2vec2, mamba, nes2net, new} have achieved strong results under these formulations. However, these approaches implicitly assume that an utterance is either entirely genuine or entirely spoofed. The ADD challenges \cite{add22, add23} and related datasets \cite{add22,add23, HAD, Psynd} have recently highlighted an issue of partial spoofing, where only specific time spans within an utterance are fake. However, even this setting does not address the scenario of component-level spoofing. Existing methods are unable to evaluate the genuineness of separate audio components in a mixture, which leads to poor performance when spoofing affects only one component of the audio scene.

To address this gap, we introduce both a new component spoofing dataset and a tailored separation-enhanced joint learning framework. The dataset contains about 2,500 utterances formed by mixing bona fide and spoofed speech or environment audio from multiple sources. Each utterance belongs to one of five categories, covering all combinations of genuine and spoofed speech and environmental sound. Building on this dataset, we propose a separation-enhanced joint learning framework: a mixture detection model first identifies utterances that may contain synthetic or substituted content, after which each part is then passed to its own anti-spoofing model: one for speech and one for environmental sound. The outputs of these models are combined and mapped to five classes. To preserve more discriminative information for spoofing detection in the speech separation module, we jointly train the separation model with the anti-spoofing models. Our contributions can be summarized as follows.
\begin{itemize}
\item We first present the component-level audio anti-spoofing concept, and introduce the first component spoofing dataset, covering diverse combinations of genuine and spoofed speech/environment audio.
\item We propose a joint learning framework that couples separation and anti-spoofing, enabling separated signals to preserve spoof-relevant information.
\item Extensive experiments demonstrate that our method outperforms the baseline, underscoring the importance of separating components and detecting spoofing for each.
\end{itemize}

\section{CompSpoof Dataset}
\label{sec:format}
The CompSpoof dataset is designed for studying component-level anti-spoofing. The dataset comprises 2,500 audio samples, evenly distributed across five classes, with 500 samples per class. Table \ref{tab:CompSpoof_classes} summarizes the class definitions.

In the mixed part of this dataset, bona fide speech comes from ASVspoof5 \cite{asvspoof5} and CommonVoice \cite{comm}, spoofed speech from ASVspoof5 \cite{asvspoof5} and SSTC \cite{sstc}, with bona fide environmental sounds from VGGSound \cite{vgg} and spoofed environmental sounds from VCapAV \cite{VCapAV}. Speech segments are chosen to contain clear voice activity, while environmental sounds are sampled from diverse scenarios such as indoor, street, and natural settings to ensure acoustic variety. In the original audio part of this dataset, we choose authentic audio utterances with both speech and simultaneously captured environmental audio signals from the VGGSound dataset \cite{vgg}. The audio durations range from 5 to 21 seconds. More details can be found at: \url{https://xuepingzhang.github.io/CompSpoof-dataset/}.

During audio processing, all files are resampled to 16 kHz, with the shorter signal determining the final duration and longer ones truncated. To control the relative prominence of speech and environmental sound, the environmental sound is adjusted in amplitude to reach a predefined SNR \cite{snr} relative to the speech.

The dataset is partitioned into training, development, and evaluation sets using stratified sampling to maintain class balance, with a ratio of 70\%, 10\%, 20\%.

\begin{table*}[t]
\caption{CompSpoof dataset class definitions}
\small
\setlength{\tabcolsep}{2.5pt} 
\begin{tabular}{ c c c c c l}
\hline
\textbf{ID} & \textbf{Mixed} & \textbf{Speech} & \textbf{Environment} & \textbf{Class Label} & \textbf{Description} \\
\hline
0 & \ding{51} & Bona fide & Bona fide &original & Original bona fide speech and corresponding environment audio without mixing \\
1 & \ding{55} & Bona fide & Bona fide &bonafide\text{\_}bonafide & Bona fide speech mixed with another bona fide environmental audio\\
2 & \ding{55} & Spoofed   & Bona fide &spoof\text{\_}bonafide & Spoof speech mixed with bona fide environmental audio \\
3 & \ding{55} & Bona fide & Spoofed   &bonafide\text{\_}spoof & Bona fide speech mixed with spoof environmental audio \\
4 & \ding{55} & Spoofed   & Spoofed   &spoof\text{\_}spoof & Spoof speech mixed with spoof environmental audio \\
\hline
\end{tabular}
\label{tab:CompSpoof_classes}

\end{table*}

\section{Separation-Enhanced Joint Learning Framework}
\subsection{Baseline}

Fig. \ref{main} a) is the baseline framework; we adopt the XLSR-AASIST model \cite{Wav2vec2}, a widely used architecture for spoofing detection. Initially designed for binary classification (bona fide vs. spoofed), we extend it to a five-class classification task corresponding to the CompSpoof dataset. Although this direct extension is straightforward, the model does not explicitly disentangle the speech and environmental components, which may lead to confusion when only one component is spoofed. The limitation motivates the introduction of a separation-based framework.

\subsection{Separation-Enhanced Joint Learning Framework}
Our method aims to explicitly separate speech and environmental sound components from an audio mixture and leverage them for robust anti-spoofing. As shown in Fig.~\ref{main}(b), the framework consists of four models: a binary mixture detection model (implemented by XLSR-AASIST \cite{Wav2vec2}), a UNet-based separation network, two dedicated anti-spoofing models for speech and environmental sound components (both implemented by XLSR-AASIST), and a joint learning mechanism that integrates their outputs. The details are introduced below.

\textbf{UNet-based Separation Network: }
To explicitly separate speech and environmental sound, we design a UNet-based separation network that operates in the STFT domain. Given an input mixed waveform, the network first computes its STFT to obtain the complex spectrogram. The speech component is estimated by predicting a complex mask via the UNet \cite{unet}, which is then applied to the mixture spectrogram. The inverse STFT (ISTFT) reconstructs the speech waveform from the masked spectrogram.

Since environmental sounds are highly diverse, obtaining a reliable environmental sound is a challenging task. We therefore compute the environmental sound in the STFT domain using an adaptive soft-mask \cite{soft}. Firstly, the remaining residual is computed by subtracting the separated speech waveform from the mixed waveform.  Let $S(f,t)$ and $R(f,t)$ denote the magnitudes of the separated speech and residual spectrograms, respectively. We first compute a dynamic scaling factor $\alpha$ to balance the speech and residual magnitudes as Eq. \ref{alpha}.
\begin{equation}
    \alpha = \frac{\text{mean}(|R(f,t)|)}{\text{mean}(|S(f,t)|) + \epsilon},
\label{alpha}
\end{equation}
where $\epsilon$ is a small constant for numerical stability. The environmental sound mask $M_\mathrm{env}(f,t)$ is then defined as Eq. \ref{env}
\begin{equation}
M_\mathrm{env}(f,t) = 1 - \tanh\Big( \frac{|S(f,t)|}{|R(f,t)| + \epsilon} \cdot \alpha \Big),
\label{env}
\end{equation}
The soft-mask serves to suppress speech leakage in the residual, preventing residual speech from being misclassified as environmental sound. Moreover, the separated environment waveform is obtained via ISTFT. Finally, the network is trained using the Mean Squared Error (MSE) between the separated and reference waveforms for both speech and environmental sound.

\textbf{Joint Learning: }
Training the separation and anti-spoofing models independently may cause the separation network to discard information that is important for detecting spoofed components. To address this, we adopt a joint learning strategy, where the separation network and the anti-spoofing models are trained together. Joint learning ensures that the separated signals retain features relevant to anti-spoofing.

In our framework, after obtaining the separated speech $W_\mathrm{sepa}^\mathrm{speech}$ and environmental sound $W_\mathrm{sepa}^\mathrm{env}$. Both the separated and the reference waveforms are then fed into the corresponding anti-spoofing models. The outputs from the separated components are compared to their target labels using cross-entropy loss $L_\mathrm{cls}^\mathrm{speech}$ and $L_\mathrm{cls}^\mathrm{env}$. In addition, a consistency loss $L_\mathrm{cons}$ is computed as the KL-divergence \cite{kl} between the predictions on the separated components and those on the reference waveform, encouraging the anti-spoofing outputs to be coherent, as shown in Eq. \ref{cons}.
\begin{equation}
\begin{aligned}
    L_\mathrm{cons} &= L_\mathrm{cons}^{env}+L_\mathrm{cons}^{speech} \\ &= \mathrm{KL}(p_\mathrm{ref}^\mathrm{env} \,\|\, p_\mathrm{sepa}^\mathrm{env}) + \mathrm{KL}(p_\mathrm{ref}^\mathrm{speech} \,\|\, p_\mathrm{sepa}^\mathrm{speech}),
\label{cons}
\end{aligned}
\end{equation}
where $p_\mathrm{ref}^\mathrm{speech}$ and $p_\mathrm{ref}^\mathrm{env}$ are the softmax outputs from the reference components in the original mixture, and $p_\mathrm{sepa}^\mathrm{speech}$ and $p_\mathrm{sepa}^\mathrm{env}$ are from the separated speech and environmental sound.

Finally, the overall joint loss $L_\mathrm{joint}$ combines the MSE separation loss $L_\mathrm{sepa}$, the mixture detection loss $L_\mathrm{cls}^\mathrm{mixed}$, the component-wise classification losses $L_\mathrm{cls}^\mathrm{speech}$ and $L_\mathrm{cls}^\mathrm{env}$, and the consistency loss $L_\mathrm{cons}$ as shown in Eq. \ref{all}
\begin{equation}
    L_\mathrm{joint} = \kappa * L_\mathrm{sepa} + L_\mathrm{cls}^\mathrm{mixed} + L_\mathrm{cls}^\mathrm{speech} + L_\mathrm{cls}^\mathrm{env} + L_\mathrm{cons},
\label{all}
\end{equation}
where $\kappa$ is a constant. Joint training ensures the separation preserves spoof-relevant features while anti-spoofing models learn from both separated components and the reference waveform.

\textbf{Inference: }During inference, the mixed waveform is first passed through the mixture detection model to obtain a binary decision (c0 vs c1234). The separation model then processes the signal to generate speech and background components, which are individually evaluated by the speech detector and environment detector, yielding their own binary decisions (c13 vs c24 and c12 vs c34). These three decisions are then combined and mapped to one of the five target classes. The above procedure produces segment-level predictions. Segment-level predictions for all chunks of an audio file are combined using majority voting to determine the final file-level label.

\begin{table}[]
\caption{Classification performance (Precision / Recall / F1) for baseline, Separation-Enhanced Framework (SEF), and Separation-Enhanced Framework with Joint Learning (SEF+JL) on dev and eval sets. The “Class” column shows IDs; the specific categories and their descriptions are provided in Table \ref{tab:CompSpoof_classes}.}
\setlength{\tabcolsep}{5pt}
\small
\centering
\begin{tabular}{ >{\centering\arraybackslash}p{1.1cm}| @{} >{\centering\arraybackslash}p{0.7cm}| c | c }
\hline
\textbf{Method}                                                                     & \,\textbf{Class} & \textbf{Dev}                            & \textbf{eval}                           \\ \hline
\multirow{6}{*}{\textbf{Baseline}}                                                   & \,0     & 1.000 / 1.000 / 1.000          & 0.962 / 1.000 / 0.980          \\
                                                                            & \,1     & 0.746 / 0.820 / 0.781          & 0.827 / 0.860 / 0.843          \\
                                                                            & \,2     & 0.811 / 0.860 / 0.835          & 0.705 / 0.790 / 0.745          \\
                                                                            & \,3     & 0.778 / 0.700 / 0.737          & 0.860 / 0.800 / 0.829          \\
                                                                            & \,4     & 0.872 / 0.820 / 0.845          & 0.793 / 0.690 / 0.738          \\
                                                                            & \;ALL   & 0.841 / 0.840 / 0.840          & 0.829 / 0.828 / 0.827            \\ \hline
\multirow{6}{*}{\textbf{SEF}}                                                        & \,0     & 1.000 / 1.000 / 1.000          & 0.990 / 1.000 / 0.995          \\
                                                                            & \,1     & 1.000 / 0.340 / 0.508          & 0.825 / 0.330 / 0.471          \\
                                                                            & \,2     & 0.710 / 0.440 / 0.543          & 0.646 / 0.420 / 0.509          \\
                                                                            & \,3     & 0.610 / 0.940 / 0.740          & 0.588 / 0.800 / 0.678          \\
                                                                            & \,4     & 0.613 / 0.920 / 0.736          & 0.561 / 0.889 / 0.688          \\
                                                                            & \;ALL   & 0.787 / 0.728 / 0.705          & 0.722 / 0.688 / 0.668          \\ \hline
\multirow{6}{*}{\begin{tabular}[c]{@{}c@{}}\textbf{SEF}\\   \;\;  \textbf{+JL}\end{tabular}} & \,0     & 1.000 / 1.000 / 1.000          & 0.980 / 1.000 / 0.990          \\
                                                                            & \,1     & 0.894 / 0.840 / 0.866          & 0.908 / 0.890 / 0.899          \\
                                                                            & \,2     & 0.860 / 0.980 / 0.916          & 0.903 / 0.840 / 0.871           \\
                                                                            & \,3     & 0.849 / 0.900 / 0.874          & 0.909 / 0.900 / 0.905          \\
                                                                            & \,4     & 0.977 / 0.840 / 0.903          & 0.841 / 0.909 / 0.874          \\
                                                                            & \;ALL   & \textbf{0.916 / 0.912 / 0.912} & \textbf{0.908 / 0.907 / 0.908} \\ \hline
\end{tabular}
\label{file_l}
\end{table}

\section{Experimental Results}
\subsection{Experimental Setup}
\textbf{Preprocessing: }
All speech samples are normalized before being fed into the models. For the baseline methods, audio preprocessing follows the same procedure as in \cite{Wav2vec2}. For the separation-based methods, audio is chunked with a window size of 4 seconds and a hop of 2 seconds. We performed Short-Time Fourier Transform (STFT) on the 16 kHz audio with a hop length of 16 ms and a window size of 64 ms. 

\textbf{Training: }
All models are trained on the same training and validation splits using Adam \cite{adam}, with learning rates of $1\times10^{-3}$ for the separation model and $1\times10^{-5}$ for the anti-spoofing models. In the joint framework, models are trained independently for the first 4 epochs, then jointly from epoch 5. In the $L_\mathrm{total}$, $\kappa=10$.

\textbf{Evaluation: }
We evaluate both the separation-based and baseline models on the dev and eval sets of CompSpoof, using file-level Precision, Recall, and F1 as the evaluation metrics. 

\subsection{Experimental results and analysis}

\textbf{Comparative experiments: }
Table \ref{file_l} shows that SEF+JL consistently outperforms both the baseline and SEF, especially for mixed-content classes where speech and environment components differ. For example, in the spoof\_bonafide class (Class ID = 2), F1 increases from 0.835 (baseline) to 0.916 (SEF+JL) on the development set. The improvement reflects that joint learning stabilizes classification across components and enhances robustness in challenging conditions.

\begin{table}[]
\centering
\caption{Segment-level detection performance (Precision / Recall / F1) on separated audio with/without Joint Learning (JL) on CompSpoof eval set.}
\begin{tabular}{c|c|c}
\hline
\textbf{Model}                                                                                 & \textbf{JL}            & \textbf{Precision / Recall / F1} \\ \hline
\multirow{2}{*}{\textbf{\begin{tabular}[c]{@{}c@{}}Speech \\ Anti-spoofing\end{tabular}}}      & \ding{51} & 0.860 / 0.875 / 0.863            \\
                                                                                               & \ding{55} & 0.777 / 0.764 / 0.720            \\ \hline
\multirow{2}{*}{\textbf{\begin{tabular}[c]{@{}c@{}}Environment \\ Anti-spoofing\end{tabular}}} & \ding{51} & 0.846 / 0.863 / 0.849            \\
                                                                                               & \ding{55} & 0.732 / 0.742 / 0.718            \\ \hline
\end{tabular}
\label{com}
\end{table}
In contrast, SEF without joint learning exhibits significant instability. While perfect performance is achieved on simple bona fide (Class ID = 0), F1 scores for mixed audio (Class ID = 1, 2, 3, 4) can drop to 0.508 or lower, indicating that separation alone may distort downstream representations without the guidance of joint learning. This emphasizes that joint optimization is critical for effectively leveraging separation in anti-spoofing detection.

\textbf{Segment-Level Model Analysis: }
Table \ref{com} presents the segment-level performance of each detection model on both separated and original signals, with and without Joint Learning (JL). Here, Segment-level metrics reflect predictions on individual audio chunks prior to aggregation.

Table \ref{com} shows that joint learning significantly improves the performance of anti-spoofing. For the speech anti-spoofing, F1 rises from 0.720 to 0.863, and for the environment anti-spoofing, F1 increases from 0.718 to 0.849. These improvements indicate that joint learning enhances the quality of separated representations and provides better supervision for downstream classification.
Environment anti-spoofing consistently performs worse than speech anti-spoofing, indicating that the XLSR-AASIST-based environment anti-spoofing model may not be suited for this task.

\section{Conclusions}
We presented a component-level audio anti-spoofing method, tackling audio component manipulations where only speech or environmental sound is forged. To support this, we constructed CompSpoof, the first dataset covering all combinations of component-wise bona fide and spoofed speech or environmental sound. We proposed a separation-enhanced joint learning framework that separates audio components and applies dedicated anti-spoofing models while preserving spoof-relevant information. Experimental results demonstrate that our method outperforms baselines, highlighting the effectiveness of component separation and joint learning.
\vfill\pagebreak
\section{Acknowledgment}
This research is funded by DKU foundation project "Emerging AI Technologies for Natural Language Processing". Many thanks for the computational resource provided by the Advanced Computing East China Sub-Center.

\small

\bibliographystyle{IEEEbib}
\bibliography{refs}

\end{document}